\documentclass [12pt]{article}
\usepackage{cite}
\begin{document}

\begin{center}
{\Large{\bf Boundary Super-Deformations, Boundary States and Tachyon
Condensation}}

\vskip .5cm {\large Zahra Rezaei} \vskip .1cm {\it Physics
Department, Tafresh University\\
P.O.Box: 39518-79611, Tafresh, Iran}\\
{\sl e-mail: z.rezaei@aut.ac.ir}\\
\end{center}

\begin{abstract}
The open string tachyon and $U(1)$ gauge field as longitudinal
fluctuations and the velocity as transverse fluctuation of an
arbitrary dimensional $D$-brane are considered as boundary
deformations of a closed superstring free action. The path integral
approach will be applied to calculate the corresponding generalized
boundary states using supersymmetrized boundary actions. Obtaining
the disk partition functions from the boundary states and studying
the effect of tachyon condensation on both of them in the NSNS and
RR sectors, leads to results that differ from the established ones.
\end{abstract}



\newpage

\section{Introduction}
$D$-branes, as unavoidable objects of string theory, can be studied
through two different approaches. On one hand, we can regard
$D$-branes as open strings boundaries because open strings are
quantum excitations of $D$-branes \cite{Uesugi,Sen1}. On the other
hand, since the boundary itself shows the creation out of vacuum
\cite{Callan1}, we can provide any $D$-brane with a boundary state
that represents the closed string creation and shows the coupling of
all closed string states to the $D$-brane \cite{Cremmer}.

The $U(1)$ gauge field (photon) and tachyon are two important states
in open string spectrum so that the former appears because of the
ending of open strings on the $D$-brane, and the latter points out
the instability of the $D$-brane. These are in fact fluctuations
along the $D$-brane world volume while the $D$-brane itself as a
dynamic object can be influenced by transverse fluctuations, too.
These transverse fluctuations, which are equivalent to taking into
account scalar fields from the world sheet point of view
\cite{Callan2}, can be interpreted as $D$-brane velocity. Boundary
states corresponding to each of these deformations (longitudinal and
transverse) have been investigated, separately in different papers
\cite{Callan1, Callan3, Billo, Lifstchytz,Divecchia, Sheikh-Jabbari,
Rezaei, Kitao, Kamani, Arfaei, Lee1, Sen2, Akhmedov}. But our main
task in this article is taking into account these longitudinal and
transverse fluctuations simultaneously as supersymmetrized
deformations of the original theory and obtaining more generalized
boundary states by the path integral method.

Actually open strings ending on bosonic and non-BPS $D$-branes, and
also stretched between $D\bar{D}$-branes contain tachyon that makes
these systems unstable. Because of the tachyon influence, an
unstable $D$-brane decays to lower dimensional configurations, and
this process is called tachyon condensation
\cite{Kutasov,Lee2,Hashimoto,Lerda}. During this process the
negative energy density of the tachyon potential at its minimum
point, cancels the tension of the $D$-brane (or $D$-branes)
\cite{Sen3}, and the final product is a closed string vacuum without
a $D$-brane or stable lower dimensional $D$-branes
\cite{Hindmarsh,Kutasov2}.

Studying tachyon condensation is possible via two main tools, open
string field theory \cite{Kostelecky,Sen4,Moeller} and boundary
string field theory
\cite{Kutasov,Witten,Shatashvili,Gerasimov,Kraus,Takayanagi,Baumgartl}.
The discussion about tachyon condensation using boundary state
formalism, which is our approach in this article, is closely related
to the latter because the boundary state normalization factor
corresponds to disk partition function which is the main component
of the latter approach.

In our previous paper \cite{Rezaei} we have also considered at the
same time the presence of the $U(1)$ gauge field, the tachyon field,
and the velocity of the $Dp$-brane, but the procedure of calculating
the boundary state was completely different. Besides, in
\cite{Rezaei} the main goal was calculating the cylindrical
amplitude between two $Dp_{1}-Dp_{2}$-branes while in this article
we are interested in disk partition functions and the effect of
tachyon condensation on them.

So in this article we consider a $U(1)$ gauge field and tachyon both
living on a $Dp$-brane world volume with arbitrary dimension. Then,
we let the $Dp$-brane have velocity along normal directions to its
world volume. Each one of these longitudinal and transverse
fluctuations will be added as a boundary action to the free action
of the theory. Then, we will introduce superfields, bosonic and
fermionic boundary coordinates, and boundary superderivatives to
find the supersymmetrized form of the mentioned deformations, which
is one of the goals of this article. Having boundary actions,
bosonic and fermionic boundary states are calculated by the path
integral approach. The profound relation between the boundary state
and the disk partition function will help us to find the NSNS and RR
partition functions. Finally, the effect of the tachyon and its
condensation on boundary states will be investigated.

Simultaneous  consideration of longitudinal and transverse
fluctuations (in spite of some technical difficulties), studied in
the framework of superstring theory and taking into account zero
modes of boundary actions and their role in the boundary state, are
the main distinctions from the conventional literature. This
generality has caused interesting deviations from standard results,
both in the boundary state and the tachyon condensation discussion,
to appear. Briefly, the disk partition function (as the
normalization factor of the boundary state) lacks the conventional
dependence on the tachyon and it causes the process of tachyon
condensation to be different. In fact, during tachyon condensation
the dimensional reduction of the $Dp$-brane occurs but the tachyon
does not completely vanish and affects the boundary state of the
newly constructed $D$-brane in the form of a constant factor.
\section{Bulk action and basis boundary states}
In order to calculate the full boundary state of a moving $Dp$-brane
in the presence of a background tachyon and U(1) gauge field, the
full sigma-model action of the closed superstring is needed. This
action can be divided into two parts, bulk and boundary. The
boundary actions will be investigated comprehensively in the next
section. Bulk action actually is the superstring free action in d=10
dimensional spacetime, and its form in terms of superfield $Y^{\mu}$
is
\begin{equation}
S=-\frac{1}{4\pi\alpha'}\int_{\Sigma}dz d\bar{z}\; d\vartheta
d\bar{\vartheta}\; g_{\mu\nu}\; DY^{\mu} \bar{D}Y^{\nu},
\end{equation}
where
\[
z=\sigma+i\tau,\;\;\; \bar{z}=\sigma-i\tau,
\]
and
\[
\vartheta=\vartheta_{+}+i\vartheta_{-},\;\;\;
\bar{\vartheta}=\vartheta_{+}-i\vartheta_{-}.
\]
$(\sigma, \tau)$ and $(\vartheta_{+}, \vartheta_{-})$ are bosonic
and Grassmann coordinates of the world sheet $\Sigma$, respectively
\cite{Bergshoeff}. $D$ and $\bar{D}$ are superderivatives that can
be shown as
\[
\left\{ \begin{array}{rcl}
D=i\frac{\partial}{\partial\bar{\vartheta}}+\frac{1}{2}\bar{\vartheta}\frac{\partial}{\partial z}\\
\bar{D}=i\frac{\partial}{\partial\vartheta}+\frac{1}{2}\vartheta\frac{\partial}{\partial
\bar{z}}.
\end{array} \right.
\]
The superfield $Y^{\mu}$ is defined in terms of spacetime
coordinates $X^{\mu}$ and their fermionic partners $\psi^{\mu}$
\begin{equation}
Y^{\mu}=X^{\mu}+\vartheta\psi_{+}^{\mu}+i\eta\bar{\vartheta}\psi_{-}^{\mu}
+i\vartheta\bar{\vartheta}B^{\mu},
\end{equation}
where $\psi_{+}^{\mu}$ and $\psi_{-}^{\mu}$ are the components of
the doublet $\psi^{\mu}$
\[\psi^{\mu}=
\left( \begin{array}{c}
\psi_{+}^{\mu} \\
\psi_{-}^{\mu}
\end{array} \right).
\]
Combining all the above relations, superstring action (1) can be
expressed in the RNS formulation as follows:
\begin{equation}
S=-\frac{1}{4\pi\alpha'}\int_{\Sigma}d^{2}\sigma
\bigg{(}g_{\mu\nu}\partial_{a}X^{\mu}\partial^{a}X^{\nu}
-ig_{\mu\nu}\bar{\psi}^{\mu}\rho^{a}\partial_{a}\psi^{\nu}\bigg{)},
\end{equation}
where $\rho$'s are Dirac matrices in two dimensions and $g_{\mu\nu}$
is constant. $X^{\mu}$ and $\psi^{\mu}$ as the solution of closed
superstring equations of motion are defined in terms of their
oscillating and zero modes as
\begin{equation}
X^{\mu}(\sigma,\tau)=x_{0}^{\mu}+2\alpha'p^\mu\tau
+\sqrt{\frac{\alpha'}{2}} \sum_{m>0}m^{-1/2}
(x^{\mu}_{m}e^{2im\sigma}+\bar{x}^{\mu}_{m} e^{-2im\sigma}),
\end{equation}
\begin{equation}
\psi_{+}^{\mu}=\sum_{m>0}\bigg{\{}\tilde{\psi}_{m}^{\mu}e^{-2im(\tau+\sigma)}
+\tilde{\psi}_{-m}^{\mu}e^{2im(\tau+\sigma)} \bigg{\}},
\end{equation}
\begin{equation}
\psi_{-}^{\mu}=\sum_{m>0}\bigg{\{}\psi_{m}^{\mu}e^{-2im(\tau-\sigma)}
+\psi_{-m}^{\mu}e^{2im(\tau-\sigma)} \bigg{\}}.
\end{equation}
In the expansion (4) $p^{\mu}$ is the closed superstring momentum
and, $x$ and $\bar{x}$ are linear combinations of the bosonic
oscillators $a$ and $\tilde{a}$
\begin{eqnarray}
\left\{
\begin{array}{rcl}
& x_{m}=a_{m}e^{-2im\tau}+\tilde{a}_{m}^{\dagger}e^{2im\tau},\\
& \bar{x}_{m}=a_{m}^{\dagger}e^{2im\tau} +\tilde{a}_{m}e^{-2im\tau},
\end{array}\right.
\end{eqnarray}
where the standard harmonic oscillators $a(\tilde{a})$ and
$a^{\dag}(\tilde{a}^{\dag})$ are related to annihilation and
creation operators $\alpha(\tilde{\alpha})$ and
$\alpha^{\dag}(\tilde{\alpha}^{\dag})$ in the following manner
\[
a_{m}^{\mu}=\frac{i}{\sqrt{m}}\alpha_{m}^{\mu},\;\;\;\;
a_{m}^{\dagger\mu}=\frac{-i}{\sqrt{m}}\alpha_{-m}^{\mu},\;\;\;\;
\tilde{a}_{m}^{\mu}=\frac{i}{\sqrt{m}}\tilde{\alpha}_{m}^{\mu},\;\;\;\;
\tilde{a}_{m}^{\dagger\mu}=\frac{-i}{\sqrt{m}}\tilde{\alpha}_{-m}^{\mu}.
\]
Also $\tilde{\psi}_{m}^{\mu}(\tilde{\psi}_{-m}^{\mu})$ and
$\psi_{m}^{\mu}(\psi_{-m}^{\mu})$ are fermionic oscillators in (5)
and (6). Defining $\theta^{\mu}$ as the boundary fermion, it should
be written as a linear combination of $\psi_{+}^{\mu}$ and
$\psi_{-}^{\mu}$
\begin{equation}
\theta^{\mu}=\psi_{+}^{\mu}+i\eta\psi_{-}^{\mu}.
\end{equation}
By considering the following oscillating form for $\theta^{\mu}$
\begin{equation}
\theta^{\mu}=\sum_{m>0}\bigg{(}\theta_{m}^{\mu}e^{-2im\sigma}+
\bar{\theta}_{m}^{\mu}e^{2im\sigma}\bigg{)},
\end{equation}
its components $\theta^{\mu}_{m}$ and $\bar{\theta}^{\mu}_{m}$ are
defined as a combination of fermionic oscillators
\begin{eqnarray}
\left\{
\begin{array}{rcl} &
\bar{\theta}_{m}^{\mu}=\tilde{\psi}_{m}^{\mu\dag}e^{2im\tau}
+i\eta\psi_{m}^{\mu}e^{-2im\tau},\\
&\theta_{m}^{\mu}=\tilde{\psi}_{m}^{\mu}e^{-2im\tau}
-i\eta\psi_{m}^{\mu\dag}e^{2im\tau} ,
\end{array}\right.
\end{eqnarray}
in analogy with (7) for the bosonic part. The relations between
fermionic oscillators are
\[
\tilde{\psi}^{\mu}_{-m}=\tilde{\psi}^{\mu\dag}_{m}\;\;\;,\;\;\;
\psi^{\mu}_{-m}=-\psi^{\mu\dag}_{m}.
\]
Equations (7) and (10) can be considered as the eigenvalue equations
\cite{Callan1}, and the corresponding eigenstates for $\tau=0$ are
\begin{equation}
|x,\bar{x}\rangle=\prod_{m=1}^\infty\exp
\bigg{(}{-\frac{1}{2}\bar{x}_{m}x_{m}-
a_{m}^{\dagger}\tilde{a}_{m}^{\dagger}+a_{m}^{\dagger}x_{m}+
\bar{x}_{m}\tilde{a}_{m}^{\dagger}}\bigg{)}|vac\rangle,
\end{equation}
as the bosonic state and
\begin{equation}
|\theta,\bar{\theta}\rangle=\prod_{m=1}^\infty\exp
\bigg{(}{-\frac{1}{2}\bar{\theta}_{m}\theta_{m}+i\eta
\psi_{m}^{\dagger}\tilde{\psi}_{m}^{\dagger}+\psi_{m}^{\dagger}\theta_{m}-i\eta
\bar{\theta}_{m}\tilde{\psi}_{m}^{\dagger}}\bigg{)}|vac\rangle,
\end{equation}
as the fermionic state. Actually, these states are basis boundary
states resulting from the action (3) which is not accompanied by any
deformations. These sets of basis boundary states can be used to
make more complicated boundary states related to nontrivial
backgrounds that couple to the original theory.

Before introducing boundary actions coupled to the original theory
we need to determine the world sheet boundary. Here we set the
boundary of the closed string world sheet at $\tau=0$ and
$\vartheta_{-}=0$ so the coordinates of this boundary are
$(\sigma,\vartheta_{+})$. Besides, tangential and normal boundary
derivatives are
\begin{equation}
\left\{
\begin{array}{rcl} &
D+\bar{D}|_{(\tau=0, \vartheta_{-}=0)}\equiv
D_{t}=i\partial_{\vartheta_{+}}+\frac{1}{2}\vartheta_{+}\partial_{\sigma},\\
& D-\bar{D}|_{(\tau=0, \vartheta_{-}=0)}\equiv
D_{n}=-\partial_{\vartheta_{-}}-\frac{i}{2}\vartheta_{+}\partial_{\tau},
\end{array}\right.
\end{equation}
which we need in the next section to couple the boundary
deformations to the theory.
\section{Boundary actions}
After writing the supersymmetrized free action, (3), to have
generalized boundary states we should add boundary actions
corresponding to boundary deformations. These boundary actions will
be supersymmetrized by construction. As mentioned before, these
deformations are open string tachyon, $U(1)$ gauge field and the
velocity of the $Dp$-brane so that the former fields are parallel to
the $Dp$-brane while the latter is normal to it. We show the
directions along the $Dp$-brane with $X^{\alpha}$ where
$\alpha\in\{0,1,...,p\}$ and the directions perpendicular to the
$Dp$-brane with $X^{i}$ so that $i\in\{p+1,...,d-1\}$. $p$ and $d$
respectively are the $Dp$-brane and spacetime dimensions. Also,
hereafter we put $\alpha'=1$ for further convenience where we want
to compare bosonic and fermionic partition functions.

\subsection{Photon}
Photons are massless particles and therefore an important part of
the open string spectrum. Since deformations related to open string
states couple to the original theory via boundary terms, the bosonic
case vector potential $A^{\mu}$ (photon) of the gauge field $U(1)$
couples to the closed string world sheet such as
$S_{F}\sim\int_{\partial\Sigma}d\sigma\;F_{\alpha\beta}
X^{\alpha}\partial_{\sigma}X^{\beta}$. In this action
$F_{\alpha\beta}$ is the field strength of $A^{\mu}$ and
$\partial_{\sigma}$ is derivative along the boundary. Also,
$\partial\Sigma$ shows the boundary of the world sheet. Since the
$U(1)$ gauge field originates from the ending of the open string on
the $Dp$-brane, so $F$ is an antisymmetric $(p+1)\times(p+1)$ matrix
with components along the $X^{\alpha}$ directions.

In analogy with the above bosonic $S_{F}$, by substituting
superfield $Y^{\mu}$ instead of $X^{\mu}$ and tangential
superderivative $D_{t}$ instead of $\partial_{\sigma}$, we can write
the supersymmetric form of $S_{F}$ as
\[
S_{F}=\frac{1}{2\pi}\int_{\partial\Sigma}d\sigma\;d\vartheta_{+}
\;F_{\alpha\beta}Y^{\alpha}D_{t}Y^{\beta}.
\]
Now we make use of the complete form of superfield $Y^{\mu}$, (2),
and perform Grassmannian integration over $\vartheta_{+}$ to find
the explicit supersymmetric form of $S_{F}$ as the follows:
\begin{equation}
S_{\rm F}=\frac{1}{2\pi}{\int}_{\partial\Sigma} d\sigma
F_{\alpha\beta}\bigg{(}X^{\alpha}\partial_{\sigma}X^{\beta}
+i\theta^{\alpha}\theta^{\beta}\bigg{)},
\end{equation}
where $\theta$ is the boundary fermion. Expansions of $X$ and
$\theta$ in terms of their oscillators help us to write the
following bosonic and fermionic forms of photon boundary action:
\begin{equation}
S_{F}^{b}=i\sum_{m>0}F_{\alpha\beta}{\bar{x}}_{m}^{\alpha}
x_{m}^{\beta},
\end{equation}
\begin{equation}
S_{F}^{f}=i\sum_{r>0}F_{\alpha\beta}{\bar{\theta}}_{r}^{\alpha}
\theta_{r}^{\beta}.
\end{equation}
Upper indices $b$ and $f$ in (15) and (16) stand for
\textit{bosonic} and \textit{fermionic}, respectively. The index $F$
indicates that the boundary action is related to the $U(1)$ gauge
field. The mode number $m$ in the bosonic part runs over the
integers while the mode number $r$ in the fermionic part runs over
the integers in the R sector and half-integers in the NS sector. In
the bosonic part there is no contribution of zero modes but since
$r$ chooses integers in the R sector of the fermionic part, there is
a zero mode contribution in the boundary action from this sector as
\begin{equation}
S_{F}^{0}=iF_{\alpha\beta}{\bar{\theta}}_{0}^{\alpha}
\theta_{0}^{\beta}.
\end{equation}

\subsection{Velocity}

To obtain the boundary state corresponding to a moving $Dp$-brane,
the boundary state in the presence of a stationary $Dp$-brane can be
obtained and then be affected by the boost operator \cite{Billo}.
However, there is another equivalent method \cite{Kamani} in which
the $Dp$-brane velocity is considered as transverse fluctuations to
the $Dp$-brane and so can be added as a boundary term to the
original action of the theory. Accordingly, the boundary action due
to the $Dp$-brane velocity in the bosonic case is $ S_{V}\sim
\int_{\partial\Sigma} d\sigma\;X^{0}V^{i}\partial_{\tau}X^{i}, $
with $V^{i}$ the $Dp$-brane velocity along the $X^{i}$ direction and
$\partial_{\tau}$ the normal derivative to the boundary.
Consequently its supersymmetric form can be written by analogy as
\[
S_{V}=\frac{1}{2\pi}\int_{\partial\Sigma}
d\sigma\;d\vartheta_{+}Y^{0}V^{i}D_{n}Y^{i},
\]
where $D_{n}$ is normal superderivative to the boundary. When we use
the complete form of $Y^{\mu}$, (2), and $D_{n}$, (13), the
supersymmetric $S_{V}$ is given by
\begin{equation}
S_{V}=\frac{1}{2\pi}{\int}_{\partial\Sigma} d\sigma
V^{i}\bigg{(}X^{0}\partial_{\tau}X^{i}
-i(\psi_{+}^{0}+i\eta\psi_{-}^{0})(\psi_{+}^{i}
-i\eta\psi_{-}^{i})\bigg{)}.
\end{equation}
Careful calculation gives the velocity action in terms of
oscillating modes in the bosonic and fermionic sectors as the
following forms:
\begin{equation}
S_{V}^{b}=\frac{i}{2}V^{i}\sum_{m>0}\bigg{\{}
\bar{x}_{m}^{0}(\tilde{a}_{m}^{i\dag}-a_{m}^{i})+
(a_{m}^{i\dag}-\tilde{a}_{m}^{i})x_{m}^{0}\bigg{\}},
\end{equation}
\begin{equation}
S_{V}^{f}=-\frac{i}{2}V^{i}\sum_{r>0}\bigg{\{}
{\bar{\theta}}_{r}^{0}(\tilde{\psi}_{r}^{i}+i\eta\psi_{r}^{i\dag})
-(\tilde{\psi}_{r}^{i\dag}-i\eta\psi_{r}^{i})
\theta_{r}^{0}\bigg{\}}.
\end{equation}
Contribution of zero modes in the velocity action limits to the
bosonic part because the equality of
$\theta_{0}^{0}=\bar{\theta}_{0}^{0}$ causes zero mode terms in the
R sector of the fermionic part to cancel each other, so
\begin{equation}
S_{V}^{0}=V^{i}x_{0}^{0}p^{i},
\end{equation}
according to (4).
\subsection{Tachyon}

Tachyon is an inevitable part of the bosonic open and closed string
spectrum. Although in superstring theories closed string tachyons
are removed by GSO projection, there are still combinations in these
theories that include tachyon. What we want to do here is consider
the open string tachyon as a deformation to original theory that
appears as a coupling to the boundary of the closed string world
sheet. This coupling in the bosonic case is $S_{T}\sim
\int_{\partial\Sigma} d\sigma T(X)$, where $T(X)$ is the tachyon
profile. In superstring theory we introduce
\begin{equation}
\Gamma^{\mu}=x^{\mu}+\vartheta\chi_{+}^{\mu}+i\eta\bar{\vartheta}\chi_{-}^{\mu}
+i\vartheta\bar{\vartheta}B^{\mu},
\end{equation}
as an auxiliary superfield in which $x^{\mu}$ and $\chi^{\mu}$ are
analogous to $X^{\mu}$ and $\psi^{\mu}$ in the main superfield
$Y^{\mu}$. By considering $T(Y)$ as a function of superfield
$Y^{\mu}$, the corresponding action can be written in the following
form:
\[
S_{\rm T}=\frac{1}{2}\int d\sigma\;d\vartheta_{+}\;(\Gamma
D_{t}\Gamma+T(Y)\Gamma),
\]
so that $D_{t}$ is the tangential derivative to the boundary. After
expanding $T(Y)$, using the $\Gamma$ and $D_{t}$ relations, (22) and
(13), and applying the Grassmannian integration over
$\vartheta_{+}$, $S_{T}$ is obtained as
\begin{equation}
S_{\rm T}=\frac{1}{2}{\int}_{\partial\Sigma} d\sigma
\bigg{(}iT^{2}+(\theta^{\mu}\partial_{\mu}T)
\partial_{\sigma}^{-1}(\theta^{\nu}\partial_{\nu}T)\bigg{)}.
\end{equation}
Since the components of the auxiliary field do not appear in the
bulk action, they can completely be eliminated by their equations of
motion. This fact has been applied to obtain (23). To have a
Gaussian integral we consider a linear profile for the tachyon, i.e.
$T^{2}(X)=X^{\mu}u_{\mu\nu}X^{\nu}$, so that $u_{\mu\nu}$ is a
constant symmetric matrix. $\partial_{\sigma}^{-1}$ actually is a
Green function that by using its following form in terms of the sign
function $\epsilon(x)$,
\[
\partial_{\sigma}^{-1}f(\sigma)=\frac{1}{2}\int
d\sigma'\;\epsilon(\sigma-\sigma')f(\sigma')\;\;\;\;,\;\;\;\
\epsilon(x)=\left\{
\begin{array}{rcl} &
-1\;\;\;\; x<0 \\
&1\;\;\;\;\;\;\; x>0,
\end{array}\right.
\]
the bosonic and fermionic parts of the tachyon action are derived in
terms of oscillators as
\begin{equation}
S_{T}^{b}=i\sum_{m>0}\frac{\pi u_{\alpha\beta}}
{2m}{\bar{x}}_{m}^{\alpha}x_{m}^{\beta},
\end{equation}
\begin{equation}
S_{T}^{f}=i\sum_{r>0}\frac{\pi u_{\alpha\beta}}
{2r}{\bar{\theta}}_{r}^{\alpha}\theta_{r}^{\beta}.
\end{equation}
Because the tachyon lives on the $Dp$-brane world volume, $u$ is a
$(p+1)\times(p+1)$ matrix that has components along the $X^{\alpha}$
directions.

Furthermore, just the bosonic part contributes in the zero mode
action as
\begin{equation}
S_{T}^{0}=\frac{i\pi}{2}u_{\alpha\beta}{\bar{x}}_{0}^{\alpha}x_{0}^{\beta}.
\end{equation}
It seems that there is a contribution of zero modes in the R sector
in which $r$ is an integer. But careful calculation of (23) for
$r=r'=0$ shows that $\int_{0}^{\pi}d\sigma
(\theta_{0}+\bar{\theta}_{0})\frac{1}{2}\int_{0}^{\pi}
d\sigma'\epsilon(\sigma-\sigma')(\theta_{0}+\bar{\theta}_{0})$ is
equal to zero. So the fermionic part has no role in the tachyon zero
mode action.
\section{Boundary States and disk partition functions}
For an arbitrary boundary action $S_{\rm boundary}$, bosonic and
fermionic boundary states are defined as
\begin{equation}
|B; S^{b}_{\rm boundary}\rangle^{\rm bosonic}=\int
D[x,\bar{x}]\;e^{iS_{\rm boundary}^{b}}\;|x,\bar{x}\rangle,
\end{equation}
\begin{equation}
|B; S^{f}_{\rm boundary}\rangle^{\rm fermionic}=\int
D[\theta,\bar{\theta}]\;e^{iS_{\rm
boundary}^{f}}\;|\theta,\bar{\theta}\rangle,
\end{equation}
where $|x,\bar{x}\rangle$ and $|\theta,\bar{\theta}\rangle$ are
basis bosonic, (11),  and fermionic, (12), boundary states,
respectively. Boundary states (27) and (28) are due to inclusion of
external background fields which present in the form of boundary
terms added to the original action. These boundary terms are called
deformations because they disturb the CFT properties of the world
sheet. $D[x,\bar{x}]$ and $D[\theta,\bar{\theta}]$ show the path
integral over $x, \bar{x}, \theta$, and $\bar{\theta}$.

To write the total boundary action we should add the boundary
actions corresponding to the boundary deformations, (15)-(17),
(19)-(21) and (24)-(26). Moreover, the bulk action itself
contributes to the boundary. The oscillating part of this
contribution helped us to form basis boundary states
$|x,\bar{x}\rangle$ and $|\theta,\bar{\theta}\rangle$, and the zero
mode part is included in the following boundary actions:
\begin{eqnarray}
S_{\rm boundary}^{b}=S_{F,T,V}^{b}=&~&Vx_{0}^{0}p^{i_{0}}+
\frac{i\pi}{2}u_{\alpha\beta}\bar{x}_{0}^{\alpha}x_{0}^{\beta}
-\frac{1}{2}g_{\mu\nu}x_{0}^{\mu}p^{\nu}\nonumber\\
&~&+i\sum_{m>0}F_{\alpha\beta}{\bar{x}}_{m}^{\alpha}
x_{m}^{\beta}+i\sum_{m>0}\frac{\pi u_{\alpha\beta}}
{2m}{\bar{x}}_{m}^{\alpha}x_{m}^{\beta}\nonumber\\
&~&+\frac{i}{2}V\sum_{m>0}\bigg{\{}\bar{x}_{m}^{0}
(\tilde{a}^{i_{0}\dag}_{m}-a_{m}^{i_{0}})+
(a_{m}^{i_{0}\dag}-\tilde{a}^{i_{0}}_{m})x_{m}^{0}\bigg{\}},
\end{eqnarray}
\begin{eqnarray}
S_{\rm boundary}^{f}=S_{F,T,V}^{f}=
&~&iF_{\alpha\beta}{\bar{\theta}}_{0}^{\alpha} \theta_{0}^{\beta}+
i\sum_{r>0}F_{\alpha\beta}{\bar{\theta}}_{r}^{\alpha}
\theta_{r}^{\beta}+i\sum_{r>0}\frac{\pi u_{\alpha\beta}}
{2r}{\bar{\theta}}_{r}^{\alpha}\theta_{r}^{\beta}\nonumber\\
&~& -\frac{i}{2}V\sum_{r>0}\bigg{\{}
{\bar{\theta}}_{r}^{0}(\tilde{\psi}_{r}^{i_{0}}+i\eta\psi_{r}^{i_{0}\dag})
-(\tilde{\psi}_{r}^{i_{0}\dag}-i\eta\psi_{r}^{i_{0}})
\theta_{r}^{0}\bigg{\}}.
\end{eqnarray}
The third term in (29) is the contribution of the bulk to the
boundary. We have supposed that the $Dp$-brane moves with the
velocity $V$ along the $X^{i_0}$ direction and the other components
of $V^i$ are zero, so the presence of $V$ and the index $i_{0}$ in
(29) and (30).

Boundary actions (29) and (30) help us to calculate the bosonic and
fermionic boundary states according to (27) and (28):
\begin{eqnarray}
|B;S_{F,T,V}^{b}\rangle^{\rm bosonic}=&~& \prod_{m=1}[\det
(-2R_{(m)})]^{-1} \exp \bigg{(}-\sum^\infty_{m=1}
a_{m}^{\dagger}\cdot S_{(m)}
\cdot\tilde{a}_{m}^{\dagger}\bigg{)}|vac\rangle
\nonumber\\
&~& \times\frac{T_p\sqrt{2^{p+1}}}{\sqrt{\det u}}\;\int
dp^{\alpha}\;\exp(-\frac{1}{2} P^T
u^{-1}P)\;\delta(x_{0}^{i_{0}}-Vx_{0}^{0}-y^{i_{0}})
\nonumber\\
&~& \times\prod_{i'\neq i_{0}}\delta(x_{0}^{i'}-y^{i'})
\prod_{\alpha}|p^{\alpha}\rangle \prod_{i'\neq i_{0}}|p^{i'}=0
\rangle|p^{i_{0}}=Vp^{0}\rangle,\\
|B;S_{F,T,V}^{f}\rangle^{\rm fermionic}=&~&
\prod_{r>0}[\det(-2R_{(r)})]
\exp\bigg{(}i\eta\sum^\infty_{r>0}\psi_{r}^{\dag}\cdot
S_{(r)}\cdot\tilde{\psi}_{r}^{\dag}\bigg{)}|vac, \theta_{0}\rangle.
\end{eqnarray}
In these boundary states, matrices $R$ and $S$ are
\begin{equation}
R_{\mu\nu}=-\frac{1}{2}g_{\mu\nu}
-F_{\alpha\beta}\delta_{\mu}^{\alpha}\delta_{\nu}^{\beta} -\frac{\pi
u_{\alpha\beta}}{2 r} \delta_{\mu}^{\alpha}\delta_{\nu}^{\beta},
\end{equation}
\begin{equation}
S_{\mu\nu(r)}=\frac{V^{2}}{4}{(R_{(r)}^{-1})}_{00}
\delta_{\mu}^{i_{0}}\delta_{\nu}^{i_{0}}
+{(R_{(r)}^{-1})}_{\alpha\beta}\delta_{\mu}^{\alpha}\delta_{\nu}^{\beta}
+g_{\mu\nu}.
\end{equation}
where $r=m$ in the bosonic case and the integer or half-integer in
the R or NS sectors of the fermionic case, respectively. The vector
$P$ in the bosonic boundary state (31) is defined in terms of the
velocity of the $Dp$-brane and the momenta of closed superstring as
\[P_{\alpha}=Vp^{i_{0}}\delta^{0}_{\;\alpha}-\frac{1}{2}p_{\alpha}.\]
The bosonic boundary state, (31), consists of two oscillating and
zero mode parts. The first line in (31) with the infinite
determinant and the exponential factor is the contribution of the
oscillators which act on the $|vac\rangle$ of oscillators. The
remaining part of (31) belongs to the zero modes with some constant
factors, two delta functions which have been included to determine
the position of the $Dp$-brane and a momentum dependent exponential
which comes from taking the zero mode action into account. By
integration over the momenta we consider the effect of all momentum
components along the $X^{\alpha}$ directions since $P^{\alpha}$ are
parallel to these directions.

Eq. (32) indicates the fermionic boundary state. Notice that the
effect of the zero mode action
$S_{F}^{0}=iF_{\alpha\beta}{\bar{\theta}}_{0}^{\alpha}
\theta_{0}^{\beta}$ on the boundary state appears as a modification
of the fermionic vacuum from $|vac\rangle$ to $|vac,
\theta_{0}\rangle$. Since our goal in this section is calculating
the disk partition function, which is obtained by projecting vacuum
onto bra-vacuum, the only state which survives is $|vac,
\theta_{0}=0\rangle$. Therefore, we do not study the explicit form
of $|vac, \theta_{0}\rangle$. In fact, when $S_{F}^{0}$ acts on the
fermionic vacuum, the polynomials of the $\Gamma$ matrices appear
which affect the spin structure of the boundary state and is
discussed in different references \cite{Callan1}.

When all the background fields and the velocity are set to zero,
$(R^{-1})_{\alpha\beta}=-2g_{\alpha\beta}$ and hence $S_{\mu\nu}$
decomposes into two parts, $S_{\alpha\beta}=-g_{\alpha\beta}$ and
$S_{ij}=g_{ij}$. Then, this boundary state belongs to a stationary
$Dp$-brane without any background fields and shows that the
directions $X^{\alpha}$, $\alpha=\{0,1,...,p\}$, and $X^{i}$,
$\alpha=\{p+1,...,d-1\}$, obey Neumann and Dirichlet boundary
conditions, respectively \cite{Divecchia}.

Since in the closed string theory, the disk partition function
represents propagation of a closed string from the boundary of the
disk and then its disappearance, so there should be a profound
relation between the boundary state and disk the partition function.
This relationship can be expressed as \cite{Arutyunov}
\begin{equation}
{\cal Z}_{S_{\rm boundary}}=\langle vac|B; S_{\rm boundary}\rangle.
\end{equation}
The index $S_{\rm boundary}$ indicates that the partition function
is corresponding to the boundary action.

Therefore, by being equipped with the generalized bosonic and
fermionic boundary states from (31) and (32), the corresponding disk
partition functions are attainable according to (35). The bosonic
disk partition function is
\begin{equation}
{\cal Z}_{\rm disk}^{b}=\frac{T_{p}\sqrt{2^{p+1}}}{\sqrt{\det
u}}\prod_{m>0}[\det(-2R_{(m)})]^{-1}\int dp^{\alpha}\exp
(-\frac{1}{2}P^{T}u^{-1}P).
\end{equation}
The partition function (36) has the factors $1/\sqrt{\det u}$ and
the infinite determinant in common with conventional partition
functions \cite{Lee1}. A significant difference is the presence of
the exponential factor of momenta, which is due to inclusion of the
zero mode parts of the boundary action [first three terms in (29)].
By performing the integration, the bosonic partition function takes
the form
\begin{equation}
{\cal Z}_{\rm disk}^{b}=\frac{T_{p}\sqrt{2^{p+1}}
\sqrt{(2\pi)^{p+1}}}{2(V^{2}-1/2)}\prod_{m>0}[\det(-2R_{(m)})]^{-1}.
\end{equation}
After the integration over the momenta it is seen that the velocity
has appeared as a coefficient and the factor $1/\sqrt{\det u}$ has
disappeared contrary to the case of a stationary $Dp$-brane
\cite{Lee1}. Actually, in the absence of velocity, just the tachyon
contributes to the zero mode boundary action and affects the
partition function by the factor $1/\sqrt{\det u}$. But with the
presence of velocity in the zero mode boundary action, the factor
$\sqrt{\det u}$ appears in the partition function which cancels the
former.

Also, in the same manner fermionic partition function is derived as
\begin{equation}
{\cal Z}_{\rm disk}^{f}=\prod_{r>0}[\det(-2R_{(r)})],
\end{equation}
in which $r$ is the integer in the RR sector and the half-integer in
the NSNS sector. As is obvious, these bosonic and fermionic
partition functions actually are the coefficients of the boundary
states (31) and (32).
\section{Tachyon condensation}
As previously mentioned, the presence of open string tachyon can be
interpreted as $D$-brane instability and shows that we have not
chosen a proper vacuum for perturbative expansion. In the other
language, since our nonlinear sigma model has broken the conformal
invariance the renormalization group (RG) flow starts from a
conformal fixed point and leaves for another fixed point. This RG
flow occurs under the influence of tachyon, which is a relevant
operator.

The tachyon potential has a minimum for $T(X)$ tending to infinity
\cite{Kutasov2} that for a profile of the form
$T^{2}(X)=u_{\mu\nu}X^{\mu}X^{\nu}$ is equivalent to
$u\rightarrow\infty$. The transition from the UV fixed point that
corresponds to the presence of an unstable $D$-brane to the IR fixed
point by tachyon condensation is accompanied by the decay of an
unstable $D$-brane to a stable vacuum or a stable $D$-brane. In the
conventional literature, the linear evolution of a single parameter
$u$ is responsible for this RG flow. Here, we have instead a
multiparameter situation that is implied by the $u_{\mu\nu}$ matrix.
Since we work in a flat spacetime, by writing down the beta
functions it will be clear that the condensation process is
independent in each coordinate. This means condensation in one
direction never stimulates condensation in the other directions. It
is different in curved backgrounds.

Endowed with the explicit form of the boundary states and partition
functions, it seems reasonable to study the effect of tachyon
condensation on them. Apart from the normalization factors of the
boundary states (31) and (32), which are partition functions, the
dependence of the boundary states on the parameters $u$ and $F$ is
summarized in the matrix $S$, (34). It is immediately clear what
happens to this matrix in the limit of infinite $u$. It will result
in $S_{(r)}\rightarrow-g$.

In order to understand this result better, suppose $F=V=0$. In this
artificial situation consider a $Dp$-brane with a single dimensional
tachyon field, $u_{pp}$, along the $X^{p}$ direction switched on, so
\[
R_{(r)\alpha\beta}=-\frac{1}{2}g_{\alpha\beta}-
\frac{u_{pp}}{2r}\delta_{\alpha}^{p}\delta_{\beta}^{p},
\]
\[
S_{(r)\mu\nu}=(R_{(r)}^{-1})_{\alpha\beta}
\delta_{\mu}^{\alpha}\delta_{\nu}^{\beta}+g_{\mu\nu},
\]
and
\[
S\rightarrow g_{ij}\oplus-g_{\alpha'\beta'}\oplus\bigg{(}
\bigg{(}-\frac{g}{2}-\frac{u}{2r}\bigg{)}^{-1}+g\bigg{)}_{pp}.
\]
We have decomposed the $\mu$ and $\nu$ indices into three parts: $i$
and $j$ show the perpendicular directions to the $Dp$-brane,
$\alpha'$ and $\beta'$ are used for directions parallel to the
$Dp$-brane world volume except $X^{p}$ and the index $p$ shows the
$X^{p}$ direction. Therefore, the matrix $S$ with and without the
influence of the tachyon will be
\[S\rightarrow
\left\{ \begin{array}{rcl} &~&
g_{ij}\oplus-g_{\alpha'\beta'}\oplus-g_{pp}\;\;\;\;\;\;\;u\rightarrow0\\
&~&
g_{ij}\oplus-g_{\alpha'\beta'}\oplus+g_{pp}\;\;\;\;\;\;\;u\rightarrow\infty.
\end{array} \right.
\]
Thus, with the change of the sign of $g_{pp}$, obviously, the
Neumann boundary condition has been changed into a Dirichlet
boundary condition. The newly generated object must therefore be a
$D(p-1)$-brane. Now if in a more general situation we consider the
tachyon field to have components along all the directions of the
$Dp$-brane world volume (i.e. $u_{\alpha\beta}$ where
$\alpha,\beta=\{0,1,...,p\}$), in the limit $u\rightarrow\infty$ all
the $p+1$ Neumann boundary conditions convert to Dirichlet boundary
conditions
\[
S\rightarrow \left\{ \begin{array}{rcl} &~&
g_{ij}\oplus-g_{\alpha\beta}\;\;\;\;\;\;\;u\rightarrow0\\
&~& g_{ij}\oplus
g_{\alpha\beta}\;\;\;\;\;\;\;\;\;\;u\rightarrow\infty.
\end{array} \right.
\]
This means that our $Dp$-brane has lost its world volume and has
reduced to a $D$-instanton. Since, the matrix $S$ has the same form
in bosonic and fermionic boundary states and the investigations show
the same results for integer and half-integer $r$, these arguments
are valid for both the bosonic and fermionic parts.

In the next step in order to complete the tachyon condensation
discussion we focus on the influence of the tachyon field on the
partition functions. As mentioned before, our tachyon background
generally has components along all the directions of the $Dp$-brane
world volume. In other words, for a $Dp$-brane, $u$ is a
$(p+1)\times(p+1)$ matrix, as $F$ is. Without loss of generality
consider $u$ as a diagonal matrix. Here, we first study the effect
of tachyon condensation on the bosonic partition function then the
method would be applied to the fermionic partition functions in the
RR and NSNS sectors, separately.

The only factor in (37) that includes the tachyon is $[\det
(-2R_{(m)})]^{-1}$. In the first step suppose that the component
$u_{pp}$ tends to infinity
\begin{eqnarray}
\lim_{u_{pp}\to\infty}\prod_{m>1}\det
\bigg{(}g_{\mu\nu}+(2F_{\alpha\beta}+\frac{u_{\alpha\beta}}{m})
\delta_{\mu}^{\alpha}\delta_{\nu}^{\beta}
\bigg{)}_{(p+1)\times(p+1)}^{-1}\nonumber\\
=\lim_{u_{pp}\to\infty}\prod_{m>1}\det
\bigg{(}g_{\mu'\nu'}+(2F_{\alpha'\beta'}+\frac{u_{\alpha'\beta'}}{m})
\delta_{\mu'}^{\alpha'}\delta_{\nu'}^{\beta'} \bigg{)}_{p\times
p}^{-1}\bigg{(}\frac{u_{pp}}{m}\bigg{)}^{-1}
\end{eqnarray}
where
\[
\mu',\nu'\in\{0,1,...,d\}-\{p\}\;\;\;and\;\;\;\alpha',\beta'\in\{0,1,...,p-1\}.
\]
This means that when $u_{pp}\to\infty$, $[\det (-2R_{(m)})]^{-1}$ is
changed to another determinant which has lost its $p$ dimension and
a factor of $(u_{pp}/m)^{-1}$ has appeared which is equal to
$\sqrt{u_{pp}}$ after regularization. So $u_{\alpha'\beta'}$ and
$F_{\alpha'\beta'}$ are symmetric and antisymmetric $p\times p$
matrices, respectively. Applying the limit $u_{pp}\to\infty$ after
regularization results in an infinite answer, which will be
discussed later. As we continue this procedure, each time by sending
any $u_{\alpha\beta}$ component to infinity, the corresponding
dimension of the determinant is reduced and a factor of
$(u_{\alpha\beta}/m)^{-1}$ presents. Finally, after a successive
process, when all the components of $u$ are tending to infinity, the
following relation is obtained
\[
\lim_{u\to\infty}\prod_{m>1}\det
\bigg{(}g_{\mu\nu}+(2F_{\alpha\beta}+\frac{u_{\alpha\beta}}{m})
\delta_{\mu}^{\alpha}\delta_{\nu}^{\beta}
\bigg{)}_{(p+1)\times(p+1)}^{-1}=\lim_{u\to\infty}\prod_{m>1}(\det
g')^{-1}\bigg{(}\det \frac{u}{m}\bigg{)}^{-1}
\]
\begin{equation}
=\lim_{u\to\infty}\sqrt{\det g'}\sqrt{(2\pi)^{p+1}\det u},
\end{equation}
where the last equality is obtained after zeta function
regularization and by $g'$ we mean the matrix of $g_{ij}$'s.

{\bf NSNS sector}

To complete the tachyon condensation process we have to study the
fermionic partition function, too. The limit $u\to\infty$ of the
fermionic partition function, (38), in the NSNS sector that $r$ is
the half-integer, leads to the following relation:
\begin{eqnarray}
\lim_{u\to\infty}\prod_{r=1/2}\det
\bigg{(}g_{\mu\nu}+(2F_{\alpha\beta}+\frac{u_{\alpha\beta}}{r})\delta_{\mu}^{\alpha}\delta_{\nu}^{\beta}
\bigg{)}_{(p+1)\times(p+1)}
\nonumber\\
=\lim_{u\to\infty}\prod_{r=1/2}(\det g')\prod_{r=1/2}\bigg{(}\det
\frac{u}{r}\bigg{)}=\lim_{u\to\infty}\sqrt{\det u},
\end{eqnarray}
where again the zeta function regularization has performed for the
last equality. Here, because $r$ is the half-integer,
$\prod_{r=1/2}(\det g')=1$. Therefore, the behavior of the total
disk partition function when $u\to\infty$ in the NSNS sector as a
combination of the bosonic, (37) and (40), and fermionic, (41),
parts is given by
\begin{equation}
{\cal Z}_{\rm disk}^{\rm
NSNS}=\lim_{u\to\infty}\frac{T_{p}(2\pi)^{p+1}
\sqrt{2^{p+1}}}{2(V^{2}-1/2)}\sqrt{\det g'}\det u.
\end{equation}

The relation between a $Dp$-brane and a $Dq$-brane tensions is
$T_{p-q}=(2\pi\sqrt{\alpha'})^{q}T_{p}$. This is correct for
$D$-branes in bosonic string theory and also BPS branes in
superstring theories. But tensions of non-BPS branes are larger by a
factor $\sqrt{2}$ with respect to BPS branes \cite{Kutasov2}. So the
relation between tensions of a BPS $Dp$-brane and a non-BPS
$D(p-1)$-brane is $\tilde T_{p-1}=2\pi \sqrt{2\alpha'}T_{p}$, where
$\tilde{T}$ shows the tension of a non-BPS brane. According to this
relation and the point that in this article $\alpha'=1$, we can read
the tension of a non-BPS $D0$-brane from (42) as
\[
\tilde{T}_{0}=(\sqrt{2})^{p}(2\pi)^{p}T_{p}.
\]
There is an interpretation for this result according to the proposed
system in this article. We can say that at the beginning there is a
non-BPS $Dp$-brane which is unstable in all $p+1$ dimensions of its
world volume due to the extension of $u$ in all of these $p+1$
directions. In the limit $u\to\infty$, condensation of the
$Dp$-brane starts in all directions of its world volume. Depending
on which components of $u$ tend to infinity, the corresponding
dimensions of the $Dp$-brane decrease. The resulting lower
dimensional $D$-brane is still non-BPS and unstable because some
components of $u$ are still available in the remaining dimensions.
When the $p$ components of $u$ tend to infinity, the $Dp$-brane
loses its $p$ spatial dimensions and a non-BPS $D0$-brane remains.
By considering the limit $u_{p+1,p+1}\to\infty$ for the last
component of $u$, the non-BPS $D0$-brane changes to a BPS
$D$-instanton with tension
\[
T_{-1}=\frac{\tilde{T}_{0} 2\pi}{\sqrt{2}}
\]
which is understandable from (42). In fact, the partition function
(42) can be written in the following form:
\begin{equation}
{\cal Z}_{\rm disk}^{\rm
NSNS}=\lim_{u\to\infty}\frac{T_{-1}}{(V^{2}-1/2)}\sqrt{\det g'}\det
u\equiv{\cal T}_{-1}.
\end{equation}
The factor $\det u$ in the partition function (43), which is absent
in conventional partition functions, stems from considering the
contribution of zero modes in boundary interactions. As an
explanation, $\det u$ can be included in $D$-instanton tension and
defines an effective tension for it, ${\cal T}_{-1}$. So the limit
$u\to\infty$ can be translated into infinite tension. In other
words, after tachyon condensation the resulting $D$-instanton has an
infinite tension that is equivalent to say that even large
interactions have no influence on the brane \cite{Zimmerman}. It
means that no higher vibration modes are excited and one expects the
brane to appear concentrated, or collapsed, in its own center of
mass \cite{Ansoldi}.

{\bf RR sector}

In the RR sector that $r$ runs over integers, $\prod_{r=1}[\det
(-2R_{(r)})]$ in the fermionic partition function cancels
$\prod_{m=1}[\det (-2R_{(m)})]^{-1}$ in the bosonic one. So the
total disk partition function in this sector is just constructed by
the zero mode part, (37), as follows:
\begin{equation}
{\cal Z}_{\rm disk}^{\rm
RR}=\frac{T_{p}\sqrt{2^{p+1}}\sqrt{(2\pi)^{p+1}}}{2(V^{2}-1/2)}.
\end{equation}
It is seen that background fields, tachyon and gauge fields, have no
contribution in this partition function. So, tachyon condensation
does not change the form of the disk partition function in the RR
sector. In other words, only the NSNS sector states are involved in
the phenomenon of tachyon condensation.

So we studied the behavior of a kind of generalized boundary state
under tachyon the condensation process. The process resembles the
conventional tachyon condensation process in decreasing the
dimension of the $Dp$-brane. But because of simultaneous
consideration of all contributions from zero modes in boundary
interactions, a tachyon dependent factor remains and defines an
effective tension for the newly generated $D$-brane.

\section{Summary and conclusion}
In this article we considered a moving and arbitrary dimensional
$D$-brane whose background fields such as open string tachyon and
$U(1)$ gauge field live on its world volume. Then we tried to couple
these nonvanishing surface terms (background fields and also
$Dp$-brane velocity) to the main action of the theory as
longitudinal and transverse boundary actions. The definition of the
boundary state in terms of boundary actions helped us to calculate
the boundary states with the path integral approach.

We divided these boundary states into zero and oscillating modes
boundary states so that each part is constructed by the
corresponding action. Inclusion of $D$-brane velocity in the problem
which is absent in conventional tachyon literature, caused our zero
mode boundary state to be different. Since we have taken into
account the zero modes of all boundary deformations as well as the
contribution of the bulk to the boundary action, the dependence of
the partition function on the tachyon differs from conventional
partition functions. In this case, during the tachyon condensation
process the phenomena of dimensional reduction of the $D$-brane and
the established relations between $D$-branes tensions occur as
expected. But a tachyon dependent factor remains in the new
partition function and defines an effective tension for the new
lower dimensional $D$-brane.

Since we have allowed the tachyon to have components along all the
directions of the $Dp$-brane world volume, condensation of all $p+1$
components of tachyon field (i.e. $u$) results in a $D$-instanton
with an effective tension that tends to infinity. This infinite
tension can be interpreted as resistance against disturbances and
fluctuations. Also it is verified that tachyon condensation is just
definable in the NSNS sector.

{\bf Acknowledgment} I would like to thank M. Baumgartl for useful
discussions. Also, I am grateful to D. Kamani for his support.


\end{document}